\documentclass[twocolumn,epjc3]{svjour3}
\usepackage{amsfonts}
\usepackage{amsmath}
\usepackage{amssymb,bbold}
\usepackage[normalem]{ulem}

\RequirePackage[T1]{fontenc}
\smartqed
\RequirePackage{graphicx}
\RequirePackage{mathptmx}
\RequirePackage{flushend}
\RequirePackage[numbers,sort&compress]{natbib}
\RequirePackage[colorlinks,citecolor=blue,urlcolor=blue,linkcolor=blue]{hyperref}
\journalname{Eur. Phys. J. C}
\RequirePackage{color}

\begin{document}

\title{Relativistic quantum dynamics of vector bosons in an Aharonov--Bohm
potential}
\author{Luis B. Castro\thanksref{e1} \and Edilberto O. Silva\thanksref{e2} }

\thankstext{e1}{e-mail: luis.castro@ufma.br}
\thankstext{e2}{e-mail: edilbertoos@pq.cnpq.br}

\institute{Departamento de F\'{\i}sica, Universidade Federal do Maranh\~{a}o, 65080-805, S\~{a}o Lu\'{\i}s, Maranh\~{a}o, Brazil.
\label{addr1}
}
\date{Received: date / Accepted: date}
\maketitle

\begin{abstract}
The Aharonov--Bohm (AB) problem for vector bosons by the
Duffin--Kemmer--Petiau (DKP) formalism is analyzed. Depending on the values of the spin projection, the relevant eigenvalue
equation coming from the DKP formalism reveals an equivalence to the spin--$
1/2$ AB problem. By using the self--adjoint extension approach, we examine
the bound state scenario. The energy spectra are explicitly computed as well
as their dependencies on the magnetic flux parameter and also the conditions
for the occurrence of bound states.

\end{abstract}

\section{Introduction}

\label{intro}

The Aharonov--Bohm (AB) effect \cite{PR.1959.115.485} has been an usual framework for
investigating the arising of phases in the wave function of quantum
particles in various physical models and has inspired a great deal of
investigations in recent years. In the AB effect, the vector potential due
to a solenoid gains an extraordinary physical meaning. It can affect the
quantum behavior of a charged particle that never encounters an
electromagnetic field. This phenomenon is intimately related to a non-local
boundary condition which relates the change in the phase of an electron wave
function to the amount of flux in the solenoid. The interest in this issue
appears in the different contexts, such as solid state physics \cite
{PRSLA.1931.130.499}, cosmic strings \cite
{PRD.2013.88.023004,Book.2000.Vilenkin,PRD.2010.82.063515,
PRD.1989.40.1346,PRL.2012.108.230404,JMP.1997.38.2553,AdP.2010.522.447,MPLA.2000.15.253,AoP.1989.193.142}
$\kappa$--Poincar\'{e}--Hopf algebra \cite{PLB.1995.359.339,PLB.2013.719.467}
, $\delta$--like singularities \cite{PRL.1990.64.503}, supersymmetry \cite
{PLB.2010.692.51}, condensed matter \cite{PRB.2011.84.125306}, Lorentz
symmetry violation \cite
{PRD.2014.90.025026,PRD.2011.83.125025,JMP.2011.52.063505}, quantum
chromodynamics \cite{PRD.2010.82.105023}, general relativity \cite
{PLB.2011.701.485}, nanophysics \cite{PRL.2010.105.036402}, quantum ring
\cite
{PRL.2012.108.106603,PRA.2010.82.022104,PLA.2011.375.2952,JMP.2012.53.023514}
, black hole \cite{PRD.2013.87.125015} and noncommutative theories \cite
{PRD.2011.84.045002,PLB.2002.527.149,EPJC.2006.46.825}.

In the AB problem of spin--$1/2$ particles a two--dimensional $\delta$--function appears as the mathematical description of the Zeeman interaction
between the spin and the magnetic flux tube \cite{PRL.1990.64.503}. This
interaction term is known to cause a splitting on the energy spectrum of
atoms depending on the spin state. In AB problem of spin--$1$ particles \cite
{PRD.1997.55.5951,PRD.1990.42.3524}, however, this characteristic is also
present. In Ref. \cite{PRD.1997.55.5951}, where the authors address the AB
problem for spin--$1$ Yang--Mills particles, it was established that, for
the case of spin--$1/2$, quasibound states exist for all noninteger flux
parameter. The existence of these states is related to the penetration of
the magnetic flux tube by the particle, which is sufficient to produce
sensitivity to the sign of the flux. The difference for spin--$1$ Yang-Mills
particles is that the quasibound states exist only for discrete values of
the magnetic flux tube, so that penetration occurs only for flux values in a
set of measure zero.

In this work, we solve the spin--$1$ AB problem for bound states in the
context of the DKP formalism. In our approach, we consider the idealized picture of a magnetic flux tube of null radius which allows the particles to access the $r=0$ region in
a controlled way. Unlike the approach taken in Ref. \cite{PRD.1997.55.5951},
here, we modulate the problem with general boundary conditions. When the spin projection $s^{3}=0$, the radial operator can be expressed as a modified Bessel differential equation. In this case, the system does not admit bound--sate solutions. On the other hand, when the spin projection $s^{1,2}=1,-1$, as we
mentioned above, we have the presence of a $\delta $--function potential in the equation of motion. As is well--known in quantum mechanics, the $\delta $--function potential guarantees at least one bound state for the particle and this property is independent of its spin. For the system considered here, in first sight, the inclusion of the spin projection element $s^{1,2}$ leads to an equation of motion equivalent to the equation for the spin--$1/2$ AB problem. Because of this, the problem can be addressed by the self--adjoint extension method \cite
{Book.1975.Reed.II,Book.2004.Albeverio} with the application of appropriate boundary conditions. After imposing the boundary conditions, we finally determine the bound states of the vector bosons in terms of the physics of the problem, in a very consistent way and without any arbitrary parameter.

\section{A short review on Duffin--Kemmer--Petiau equation}
\label{sec:1}

The first--order DKP formalism \cite
{Petiau1936,Kemmer1938,PR54:1114:1938,Kemmer1939} describes spin--$0$ and
spin--$1$ particles and has been used to analyse relativistic interactions
of spin--$0$ and spin--$1$ hadrons with nuclei as an alternative to their
conventional second--order Klein--Gordon (KG) and Proca counterparts.
Although the formalisms are equivalent in the case of minimally coupled
vector interactions \cite{PLA244:329:1998,PLA268:165:2000,PRA90:022101:2014}, the DKP formalism enjoys a richness of couplings not capable of being
expressed in the KG and Proca theories \cite{PRD15:1518:1977,JPA12:665:1979}. Indeed, the DKP formalism has been widely used in the description of many processes in elementary particle and nuclear physics and it proved to be better than the KG formalism in the analysis of $K_{l3}$
decays, the decay--rate ratio $\Gamma(\eta\rightarrow
\gamma\gamma)/\Gamma(\pi^{0}\rightarrow \gamma\gamma)$, and level shifts and
widths in pionic atoms \cite{PRL26:1200:1971,PTP51:1585:1974,PRC34:2244:1986}. The DKP formalism has also applications in other contexts, as such, in
noncommutative phase space \cite{EPJC.2012.72.2217}, in Very Special
Relativity (VSR) symmetries \cite{EPJP.2014.129.246}, in Bose--Einstein
condensates \cite{PLA316:33:2003,PA419:612:2015}, in topological defects
\cite{EPJC.2015.75.287}, in thermodynamics properties \cite{ADHEP2015:901675:2015}, in topological semimetals \cite{PRB92:235106:2015}, in noninertial effect of rotating frames \cite{EPJC76:61:2016}, among others.

The DKP equation for a free charged boson is given by \cite{Kemmer1939}
(with units in which $\hbar =c=1$)
\begin{equation}
\left( i\beta ^{\mu }\partial _{\mu }-M\right) \Psi =0,  \label{dkp}
\end{equation}
\noindent where the matrices $\beta ^{\mu }$\ satisfy the DKP algebra
\begin{equation}
\beta ^{\mu }\beta ^{\nu }\beta ^{\lambda }+\beta ^{\lambda }\beta ^{\nu
}\beta ^{\mu }=g^{\mu \nu }\beta ^{\lambda }+g^{\lambda \nu }\beta ^{\mu
}\noindent ,
\end{equation}
\noindent and the metric tensor is $g^{\mu \nu }=\,$diag$\,(1,-1,-1,-1)$.
That algebra generates a set of 126 independent matrices whose irreducible
representations are a trivial representation, a five--dimensional
representation describing the spin--$0$ particles and a ten--dimensional
representation associated to spin--$1$ particles. The DKP spinor has an
excess of components and the theory has to be supplemented by an equation
which allows to eliminate the redundant components. That constraint equation
is obtained by multiplying the DKP equation by $1-\beta ^{0}\beta ^{0}$,
namely
\begin{equation}
i\beta ^{j}\beta^{0}\beta ^{0}\partial _{j}\Psi =M\left( 1-\beta ^{0}\beta
^{0}\right) \Psi,  \label{vin1}
\end{equation}
\noindent where $j$ runs from 1 to 3. This constraint equation expresses
three (four) components of the spinor by the other two (six) components and
their space derivatives in the scalar (vector) sector so that the
superfluous components disappear and there only remain the physical
components of the DKP theory. The second--order KG and Proca equations are
obtained when one selects the spin--$0$ and spin--$1$ sectors of the DKP
theory. A well--known conserved four--current is given by
\begin{equation}
J^{\mu }=\frac{1}{2}\bar{\Psi}\beta ^{\mu }\Psi,  \label{jota}
\end{equation}
\noindent\ where the adjoint spinor $\bar{\Psi}$ is given by $\bar{\Psi}
=\Psi ^{\dagger }\eta ^{0}$ with $\eta ^{0}=2\beta^{0}\beta ^{0}-1$ in such
a way that $\left( \eta ^{0}\beta ^{\mu }\right)^{\dagger }=\eta ^{0}\beta
^{\mu }$ (the matrices $\beta ^{\mu }$ are Hermitian with respect to $\eta
^{0}$). Despite the similarity to the Dirac equation, the DKP equation
involves singular matrices, the time component of $J^{\mu }$ given by (\ref
{jota}) is not positive definite and the case of massless bosons can not be
obtained by a limiting process \cite{PRD10:4049:1974}. Nevertheless, the
matrices $\beta ^{\mu }$ plus the unit operator generate a ring consistent
with integer--spin algebra and $J^{0}$ may be interpreted as a charge
density. The factor $1/2$ multiplying $\bar{\Psi}\beta ^{\mu }\Psi $, of no
importance regarding the conservation law, is in order to hand over a charge
density conformable to that one used in the KG theory and its
nonrelativistic limit \cite{JPA43:055306:2010}.

\section{Interactions in the Duffin--Kemmer--Petiau equation}

With the introduction of interactions, the DKP equation can be written as
\begin{equation}
\left(i\beta^{\mu}\partial_{\mu }-M-U\right) \Psi =0,  \label{dkp222}
\end{equation}
where the more general potential matrix $U$ is written in terms of 25 (100)
linearly independent matrices pertinent to five (ten)--dimensional
irreducible representation associated to the scalar (vector) sector. In the
presence of interaction, $J^{\mu}$ satisfies the equation
\begin{equation}  \label{corrent222}
\partial _{\mu }J^{\mu }+\frac{i}{2}\bar{\Psi}\left( U-\eta ^{0}U^{\dagger
}\eta^{0}\right) \Psi =0.
\end{equation}
\noindent Thus, if $U$ is Hermitian with respect to $\eta ^{0}$, then
four-current will be conserved. The potential matrix $U$ can be written in
terms of well--defined Lorentz structures. For the spin--$0$ (scalar sector)
there are two scalar, being two vector and two tensor terms \cite
{PRD15:1518:1977}, whereas for the spin--$1$ (vector sector) there are two
scalar, two vector, a pseudoscalar, two pseudovector and eight tensor terms
\cite{JPA12:665:1979}. The condition (\ref{corrent222}) has been used to
point out a misleading treatment in the recent literature regarding
analytical solutions for nonminimal vector interactions \cite
{CJP87:857:2009,JPA45:075302:2012,ADHEP2014:784072:2014}.

\subsection{Duffin--Kemmer--Petiau equation with minimal electromagnetic
coupling}

Considering only the minimal vector interaction, the DKP equation for a
charged boson with minimal electromagnetic coupling is given by
\begin{equation}
\left( i\beta ^{\mu }D_{\mu }-M\right) \Psi =0,  \label{dkp2}
\end{equation}
\noindent where the covariant derivative is given by $D_{\mu }=\partial
_{\mu}+ieA_{\mu}$. In this case, the constraint equation (\ref{vin1})
becomes
\begin{equation}
i\beta ^{k}\beta ^{0}\beta ^{0}\partial _{k}\Psi -e\beta^{k}\beta ^{0}\beta
^{0}A_{k}\Psi =M\left( 1-\beta ^{0}\beta ^{0}\right) \Psi,  \label{vin2}
\end{equation}
\noindent and the four--current $J^{\mu }$ retains its form as (\ref{jota}).

\subsection{Vector sector}

Now, we discuss the vector sector (spin--$1$ sector) of the DKP theory. To
select the physical component of the DKP field for the vector sector (spin--$%
1$ sector), we define the operator \cite{UMEZAWA1956}
\begin{equation}
R^{\mu }=(\beta ^{1})^{2}(\beta ^{2})^{2}(\beta ^{3})^{2}\left[ \beta ^{\mu
}\beta ^{0}-g^{\mu 0}\right] ,  \label{oper}
\end{equation}
\noindent which satisfies $R^{\mu \nu }=R^{\mu }\beta ^{\nu }$ and $R^{\mu
\nu }=-R^{\nu \mu }$. Moreover, as it is shown in Ref. \cite{UMEZAWA1956}, $
R^{\mu }\Psi $ and $R^{\mu \nu }\Psi $ transform as a (pseudo)vector and
(pseudo)tensor quantities under an infinitesimal Lorentz transformation,
respectively. From the above definitions, the following property is
obtained:
\begin{equation}
R^{\mu \nu }\beta ^{\alpha }=R^{\mu }g^{\nu \alpha }-R^{\nu }g^{\mu \alpha
}\,.  \label{v1}
\end{equation}
\noindent In this way, by applying the $R^{\mu }$ and $R^{\mu \nu }$
operators to the DKP equation (\ref{dkp2}), we obtain
\begin{subequations}
\begin{align}
&D_{\mu }\left( R^{\nu \mu }\Psi \right) =-iM\left( R^{\nu }\Psi \right) ,
\\
&\left( R^{\mu \nu }\Psi \right) =-\frac{i}{M}U^{\mu \nu }, \\
&U^{\mu \nu }=D^{\mu }\left( R^{\nu }\Psi \right) -D^{\nu }\left( R^{\mu
}\Psi \right) ,
\end{align}
which leads to
\end{subequations}
\begin{subequations}
\label{v6}
\begin{align}
&D_{\mu }U^{\mu \nu }+M^{2}\left( R^{\nu }\Psi \right) =0, \\
&D_{\mu }\left( R^{\mu }\Psi \right) =\frac{ie}{2M^{2}}F_{\mu \nu }U^{\mu
\nu },
\end{align}
where $F_{\mu \nu }=\partial _{\mu }A_{\nu }-\partial _{\nu }A_{\mu }$.
These results tell us that all elements of the column matrix $R^{\mu }\Psi $
obey the Proca equation interacting minimally with an electromagnetic field.
So, this procedure selects the vector sector of DKP theory, making
explicitly clear that it describes a spin--$1$ particle embedded in a
electromagnetic field.

According to Ref. \cite{JPA43:495402:2010}, we can rewrite Eq. (\ref{v6}) in
the form
\end{subequations}
\begin{equation}
\left[ D_{\mu }D^{\mu }+M^{2}\right] R^{\nu }\Psi -D^{\nu }D_{\mu }R^{\mu
}\Psi -\frac{ie}{2} R^{\nu }S^{\alpha\mu }F_{\mu \alpha }\Psi =0,
\label{neweq2}
\end{equation}
\noindent where $S^{\mu \nu }=\left[ \beta ^{\mu },\beta ^{\nu }\right] $.
The term $D^{\nu}D_{\mu}R^{\mu}\Psi$ is called the anomalous term because it
has no equivalent in the spin--$1/2$ Dirac theory \cite{Kemmer1939}.
However, it has been shown in Refs. \cite{PLA244:329:1998,PLA268:165:2000}
that such an anomalous term disappears when the physical components of the
DKP field are selected.

Following the same procedure of Ref. \cite{JPA43:495402:2010}, Eq. (\ref
{neweq2}) becomes
\begin{equation}
\left[ D_{\mu }D^{\mu }+M^{2}-e\left( \mathbf{S}\cdot \mathbf{B}\right)
\right] \mathbf{R}\Psi =0.  \label{neweq3}
\end{equation}
with $\mathbf{B}=\mathbf{\nabla }\times \mathbf{A}$ and the spin operator $
\mathbf{S}=\left( S^{1},S^{2},S^{3}\right) $ is expressed by
\begin{equation}
S^{1}=
\begin{pmatrix}
0 & 0 & 0 \\
0 & 0 & -i \\
0 & i & 0
\end{pmatrix}
\,,~S^{2}=
\begin{pmatrix}
0 & 0 & i \\
0 & 0 & 0 \\
-i & 0 & 0
\end{pmatrix}
\,,~S^{3}=
\begin{pmatrix}
0 & -i & 0 \\
i & 0 & 0 \\
0 & 0 & 0
\end{pmatrix}
\,.  \label{s1}
\end{equation}
\noindent In this stage, it is worthwhile to mention that the last term in
equation (\ref{neweq3}) is the called Pauli term, which is crucial to give
meaning to the term that explicitly depends of the spin. Also, we can
mention that the Pauli term is only important for the vector sector (spin--$
1 $ sector) of the DKP theory, because that term is absent for the scalar
sector (spin--$0$ sector). For this reason, we only focus the vector sector
of the DKP theory.

If the terms in the potential $A^{\mu}=(A_{0},\mathbf{A})$ are
time--independent one can write
\begin{equation}
\Psi (\mathbf{r},t)=\psi (\mathbf{r})\mathrm{e}^{-iEt},
\end{equation}
\noindent where $E$ is the energy of the vector boson, in such a way that
the time--independent DKP equation for the vector sector becomes
\begin{equation}
\left[\left( \mathbf{p}-e\mathbf{A}\right)^{2}+M^{2}-(E-eA_{0})^{2}-e\left(
\mathbf{S}\cdot \mathbf{B}\right) \right]\mathbf{R}\psi=0.  \label{hm}
\end{equation}
\noindent In the next section, we apply Eq. (\ref{hm}) to AB problem, giving
a focus to spin effects through the $e\left(\mathbf{S}\cdot \mathbf{B}
\right) $ term. We shall see later that by carefully modulating the radial
operator (\ref{hm}) with boundary conditions, it can provide both bound and
scattering states. However, we emphasize only bound states.

\section{The Aharonov--Bohm problem}

Let us consider the particular case where the boson moves in the presence of
the AB potential $(A_{0}=0)$. The vector potential in the Coulomb gauge is
\begin{equation}
e\mathbf{A}=-\frac{\phi }{\rho }\mathbf{\hat{\varphi}},  \label{ab}
\end{equation}
\noindent where $\phi $ is the flux parameter. The potential in (\ref{ab})
provides a magnetic field perpendicular to the plane $\left( \rho ,\varphi
\right) $, namely
\begin{equation}
e\mathbf{B}=-\phi \frac{\delta (\rho )}{\rho }\mathbf{\hat{z}},  \label{abcm}
\end{equation}
\noindent where $\mathbf{B}$ is the magnetic field due to a solenoid. If the
solenoid is extremely long, the field inside is uniform, and the field
outside is zero. However, the boson is allowed to access the $\rho =0$
region. In this region, the magnetic field is non--null. If the radius of
the solenoid is $\rho _{0}\approx 0$, then the relevant magnetic field is $
B\sim \delta (\rho )$ as in (\ref{abcm}).

\subsection{Aharonov-Bohm problem for the spin--1 sector}

Now, we consider the effect of Aharonov--Bohm flux field on vector
bosons. Substituting Eq. (\ref{ab}) in Eq. (\ref{hm}), we obtain
\begin{equation}
\left[ \left( \frac{1}{i}\boldsymbol{\nabla }+\frac{\phi }{\rho }\mathbf{
\hat{\varphi}}\right) ^{2}+\phi S\frac{\delta (\rho )}{\rho }\right] \mathbf{
R}\psi =(E^{2}-M^{2})\mathbf{R}\psi \,,  \label{neweq4}
\end{equation}
\noindent where $S$ is the matrix
\begin{equation}
S=
\begin{pmatrix}
1 & 0 & 0 \\
0 & -1 & 0 \\
0 & 0 & 0
\end{pmatrix}
\,,  \label{sz}
\end{equation}
\noindent and
\begin{equation}
\boldsymbol{\nabla }=\mathbf{\hat{\rho}}\frac{\partial }{\partial \rho }+
\mathbf{\hat{\varphi}}\frac{1}{\rho }\frac{\partial }{\partial \varphi }+
\mathbf{\hat{z}}\frac{\partial }{\partial z},  \label{laplacc}
\end{equation}
\noindent is the gradient operator in cylindrical coordinates. \noindent At
this stage, we can use the invariance under boosts along the z--direction
and adopt the usual decomposition
\begin{equation}
\mathbf{R}\psi =R^{i}\psi (\rho ,\varphi ,z)=f_{m}^{(i)}(\rho )e^{im\varphi}e^{ip_{z}z},  \label{ansatz2}
\end{equation}
\noindent with $m\in \mathbb{Z}$. Inserting Eq. (\ref{ansatz2}) into Eq. (
\ref{neweq4}), we get
\begin{equation}\label{dra}
\mathfrak{O}f_{m}^{(i)}\left( \rho \right) =k^{2}f_{m}^{(i)}\left( \rho
\right) ,
\end{equation}
\noindent with $k=\sqrt{E^{2}-M^{2}-p_{z}^{2}}$, where
\begin{equation}
\mathfrak{O}=\mathfrak{O}_{0}+\phi s^{i}\frac{\delta (\rho )}{\rho },
\label{hamiltonian}
\end{equation}
\begin{equation}
\mathfrak{O}_{0}=-\frac{d^{2}}{d\rho ^{2}}-\frac{1}{\rho }\frac{d}{d\rho }+
\frac{(m+\phi )^{2}}{\rho ^{2}},  \label{freeH}
\end{equation}
\noindent and $s^{i}=\left( 1,-1,0\right) $ represents the eigenvalues of
the operator $S$ acting on the DKP spinor $f_{m}^{(i)}(\rho )$. Equation (
\ref{dra}) describes the quantum dynamics of vector bosons in the presence
of the Aharonov--Bohm potential. From (\ref{dra}) we can see that scattering states occur only if $k\in\mathbb{R}$, whereas bound states occur only if $k=i|k|$.

At this level, it is worthwhile to note that the solution for
this problem can be separated in two cases. The first case is when the spin projection
$s^{3}=0$. In this case the Pauli term is absent and the radial operator
$\mathfrak{O}$ becomes $\mathfrak{O}_{0}$ and $f^{(3)}_{m}$ can be expressed as a solution of the modified Bessel differential equation. We can see that the system does not admit bound--state solutions. On the other hand, the second case is when $s^{1,2}=1,-1$. In this case the radial operator $\mathfrak{O}$ is equivalent to Eq. (29)
of Ref. \cite{AoP.2013.339.510} (see also Refs. \cite
{PRL.1990.64.503,PRD.1994.50.7715,PRD.2012.85.041701}) which governs the
quantum dynamics of the usual spin--$1/2$ AB problem, namely
\begin{equation}
-\frac{d^{2}}{dr^{2}}-\frac{1}{r}\frac{d}{dr}+\frac{(m+\phi )^{2}}{r^{2}}
+\phi s\frac{\delta (r)}{r},  \label{Hab}
\end{equation}
\noindent where $s$ is twice the spin value, with $s=+1$ for spin
\textquotedblleft up\textquotedblright\ and $s=-1$ for spin
\textquotedblleft down\textquotedblright , and $\phi $ is the flux parameter.

In order to study the dynamics of the system for
$s^{1,2}=1,-1$, it is necessary to solve Eq. (\ref{dra}). However, as for
the case studied in Ref. \cite{AoP.2013.339.510}, it also involves a
singularity in the $r=0$ region. The appropriate method for studying this
problem is the self--adjoint extension method of operators in quantum
mechanics. By exploiting the nature of the $\delta $--function in Eq. (\ref
{hamiltonian}), we can see that the system admits at least a bound state. In
addition, we also must verify that the operator $\mathfrak{O}_{0}$ in Eq. (
\ref{freeH}) admits or not self--adjoint extensions. It is known that when
we consider Eq. (\ref{hamiltonian}), that is, taking into account the $
\delta $--function, $\mathfrak{O}_{0}$ is not essentially self--adjoint. In
this case, we must find the self--adjoint extensions of the operator $
\mathfrak{O}_{0}$ corresponding to different types of boundary conditions.
Such self--adjoint extensions are based in boundary conditions at the origin
and conditions at infinity \cite
{crll.1987.380.87,JMP.1998.39.47,LMP.1998.43.43}. From the theory of
symmetric operators, it is a well--known fact that the symmetric radial
operator $\mathfrak{O}_{0}$ is essentially self--adjoint if
\begin{equation}
\left\vert m+\phi \right\vert \geq 1,  \label{ess}
\end{equation}
\noindent while for
\begin{equation}
\left\vert m+\phi \right\vert <1,  \label{saex}
\end{equation}
\noindent it admits an one-parameter family of self--adjoint extensions \cite
{Book.1975.Reed.II}, $\mathfrak{O}_{0,\zeta _{m}^{i}}$, where $\zeta _{m}^{i}
$ is the self--adjoint extension parameter. According to Ref. \cite
{AoP.2013.339.510}, the operator (\ref{freeH}) admits an one--parameter
family of self--adjoint extensions. To characterize this parameter family of
self--adjoint extension, we use the approaches proposed by Kay--Studer (KS)
\cite{CMP.1991.139.103} and Bulla--Gesztesy (BG) \cite{JMP.1985.26.2520},
being both based on boundary conditions. In short, in the KS approach, the
boundary condition is a match of the logarithmic derivatives of the
zero--energy solutions for Eq. (\ref{dra}) and the solutions for the problem
$\mathfrak{O}_{0}$ plus self--adjoint extension. In the BG approach,
however, the boundary condition is a mathematical limit allowing divergent
solutions for the operator $\mathfrak{O}_{0}$ in Eq. (\ref{freeH}) at
isolated points, provided they remain square integrable. Then, following
Ref. \cite{AoP.2013.339.510}, the energy spectrum using the KS approach is
found to be
\begin{align}
\left( E_{m}^{i}\right)^{2} &=M^{2}-\frac{4}{\rho _{0}^{2}}\left[ \left(
\frac{\phi s^{i}+\left\vert m+\phi \right\vert }{\phi s^{i}-\left\vert
m+\phi \right\vert }\right) \frac{\Gamma (1+\left\vert m+\phi \right\vert )}{
\Gamma (1-\left\vert m+\phi \right\vert )}\right] ^{\frac{1}{\left\vert
m+\phi \right\vert }}  \notag \\
& +p_{z}^{2},  \label{energyKS}
\end{align}
where $\rho _{0}$ is a finite very small radius. This radius may be
understood as a kind of physical regularization for the $\delta $--function
in Eq. (\ref{hamiltonian}). Moreover, according to the BG method, the energy
spectrum is given by
\begin{equation}
\left( E_{m}^{i}\right) ^{2}-M^{2}=-4\left( -\frac{1}{\zeta _{m}^{i}}\frac{
\Gamma \left( 1+\left\vert m+\phi \right\vert \right) }{\Gamma \left(
1-\left\vert m+\phi \right\vert \right) }\right) ^{\frac{1}{\left\vert
m+\phi \right\vert }}+p_{z}^{2}.  \label{energyBG}
\end{equation}%
We can note in Eq. (\ref{energyKS}) that the KS method gives us energy
levels without any arbitrary parameter that can come from it, while the BG
method gives an expression that leaves an arbitrary parameter, namely, the
self--adjoint extension of the parameter $\zeta _{m}^{i}$. However, if we
directly compare Eqs. (\ref{energyKS}) and (\ref{energyBG}), an expression
for the self--adjoint extension parameter $\zeta _{m}^{i}$ is exactly found,
i.e.,
\begin{equation}\label{zetaj}
\zeta _{m}=-\rho _{0}^{2\left\vert m+\phi \right\vert }\left( \frac{\phi
s^{i}-\left\vert m+\phi \right\vert }{\phi s^{i}+\left\vert m+\phi
\right\vert }\right) .
\end{equation}
\noindent The component of the DKP spinor for bound-state solutions for the spin projection $s^{1,2}=1,-1$ is given by
\begin{equation}
R^{i}\psi (\rho ,\varphi ,z)=C_{m}\;K_{\left\vert m+\phi \right\vert }\left(
|k^{(i)}|\rho\right) e^{im\varphi }e^{ip_{z}z},
\end{equation}
where $C_{m}$ is a normalization constant, $K_{\left\vert m+\phi \right\vert
}$ are the modified Bessel functions of second kind and $|k^{(i)}|=\sqrt{M^{2}+p_{z}^{2}-\left(
E_{m}^{i}\right) ^{2}}$ can be obtained from Eq. (\ref{energyKS}). This important result was found for the first time in Ref. \cite
{PRD.2012.85.041701}, where was proposed a general regularization procedure
to obtain the self--adjoint extension parameter for both state bound and
scattering problem for the spin--$1/2$ AB problem in conical space in $(1+2)$
dimensions. To ensure that Eq. (\ref{energyBG}) is a real number, the
self--adjoint extension parameter must be negative, i.e., $\zeta _{m}^{i}<0$
. This condition, however, ensures that the system admits relativistic bound
states.

\section{Conclusions}
\label{sec:4}

In this work, we have addressed the spin--1 AB problem in the context of the
DKP formalism. We have assumed vector bosons incidents on a flux tube,
characterized by a $\delta$--function. We found that our problem can be separated in two cases depending on the spin projection. For $s^{3}=0$, the Pauli's term is absent ($\delta$--function is absent), the radial operator can be expressed as a modified Bessel differential equation and we can see that the system does not admit bound--state solutions. Otherwise, for $s^{1,2}=1,-1$, the radial operator is equivalent to usual spin--$1/2$ AB problem and consequently, equivalent to the problem of a particle in the presence of a $\delta$--function potential in one dimension in quantum mechanics. The easiest way
of dealing with singularities in quantum mechanics is by imposing that the
eigenfunction vanishes at the singularity. However, although convenient,
this does not necessarily give the best description (or even the correct
one) of the physical phenomenon studied. Care must be taken to insure that
the operator is self--adjoint in the region of interest. This can be
achieved by extending the domain of the operator $\mathfrak{O}_{0}$ in Eq. (\ref{freeH}) to equal that of $\mathfrak{O}_{0}^{\dag }$. By doing this, a
family of boundary conditions might appear, including the one that requires
the eigenfunction to vanish at the singularity. With this in hands, physics
itself determines the appropriate boundary condition. In this sense, the
self--adjoint extension approach was used to determine the bound states of
vector bosons for the spin projection $s^{1,2}=1,-1$ in terms of the physics of the problem, in a very consistent
way and without any arbitrary parameter. Finally, expressions for the bound
states energy for vector bosons in the presence of the AB potential has been
obtained and the conditions in which they occur were established.

\begin{acknowledgements}
This work was supported by the CNPq, Brazil, Grants No. 482015/2013--6
(Universal), No 455719/2014--4 (Universal), No. 306068/2013--3 (PQ),
304105/2014--7 (PQ) and FAPEMA, Brazil, Grants No. 01852/14 (PRONEM).
\end{acknowledgements}

\bibliographystyle{spphys}
\bibliography{mybibfile_stars2}

\begin{thebibliography}{10}
\providecommand{\url}[1]{{#1}}
\providecommand{\urlprefix}{URL }
\expandafter\ifx\csname urlstyle\endcsname\relax
  \providecommand{\doi}[1]{DOI \discretionary{}{}{}#1}\else
  \providecommand{\doi}{DOI \discretionary{}{}{}\begingroup
  \urlstyle{rm}\Url}\fi

\bibitem{PR.1959.115.485}
Y.~Aharonov, D.~Bohm, Phys. Rev. \textbf{115}(3), 485 (1959).
\newblock \doi{10.1103/PhysRev.115.485}

\bibitem{PRSLA.1931.130.499}
R.L. Kronig, W.G. Penney, Proc. R. Soc. Lond. A \textbf{130}, 499 (1931).
\newblock \doi{10.1098/RSPA.1931.0019}

\bibitem{PRD.2013.88.023004}
M.~Nouri-Zonoz, A.~Parvizi, Phys. Rev. D \textbf{88}(2), 023004 (2013).
\newblock \doi{10.1103/PhysRevD.88.023004}

\bibitem{Book.2000.Vilenkin}
A.~Vilenkin, E.P.S. Shellard, \emph{Cosmic Strings and Other Topological
  Defects} (Cambridge University Pres, Canbridge, 2000)

\bibitem{PRD.2010.82.063515}
Y.Z. Chu, H.~Mathur, T.~Vachaspati, Phys. Rev. D \textbf{82}(6), 063515 (2010).
\newblock \doi{10.1103/PhysRevD.82.063515}

\bibitem{PRD.1989.40.1346}
P.~de~Sousa~Gerbert, Phys. Rev. D \textbf{40}(4), 1346 (1989).
\newblock \doi{10.1103/PhysRevD.40.1346}

\bibitem{PRL.2012.108.230404}
M.A. Hohensee, B.~Estey, P.~Hamilton, A.~Zeilinger, H.~M\"{u}ller, Phys. Rev.
  Lett. \textbf{108}(23), 230404 (2012).
\newblock \doi{10.1103/PhysRevLett.108.230404}

\bibitem{JMP.1997.38.2553}
V.B. Bezerra, J. Math. Phys. \textbf{38}(5), 2553 (1997).
\newblock \doi{10.1063/1.531995}

\bibitem{AdP.2010.522.447}
K.~Bakke, C.~Furtado, Ann. Phys. (Berlin) \textbf{522}(7), 447 (2010).
\newblock \doi{10.1002/andp.201000043}

\bibitem{MPLA.2000.15.253}
C.~Furtado, V.~B.~Bazerra, F.~Moraes, Mod. Phys. Lett. A \textbf{15}(4), 253
  (2000).
\newblock \doi{10.1142/S0217732300000244}

\bibitem{AoP.1989.193.142}
A.~Aliev, D.~Gal'tsov, Ann. Phys. \textbf{193}(1), 142 (1989).
\newblock \doi{10.1016/0003-4916(89)90355-2}

\bibitem{PLB.1995.359.339}
P.~Roy, R.~Roychoudhury, Phys. Lett. B \textbf{359}(3–4), 339 (1995).
\newblock \doi{10.1016/0370-2693(95)01079-6}

\bibitem{PLB.2013.719.467}
F.M. Andrade, E.O. Silva, Phys. Lett. B \textbf{719}(4-5), 467 (2013).
\newblock \doi{10.1016/j.physletb.2013.01.062}

\bibitem{PRL.1990.64.503}
C.R. Hagen, Phys. Rev. Lett. \textbf{64}(5), 503 (1990).
\newblock \doi{10.1103/PhysRevLett.64.503}

\bibitem{PLB.2010.692.51}
V.~Jakubsk\'{y}, L.M. Nieto, M.S. Plyushchay, Phys. Lett. B \textbf{692}(1), 51
  (2010).
\newblock \doi{10.1016/j.physletb.2010.07.014}

\bibitem{PRB.2011.84.125306}
A.O. Slobodeniuk, S.G. Sharapov, V.M. Loktev, Phys. Rev. B \textbf{84}, 125306
  (2011).
\newblock \doi{10.1103/PhysRevB.84.125306}

\bibitem{PRD.2014.90.025026}
H.~Belich, K.~Bakke, Phys. Rev. D \textbf{90}, 025026 (2014).
\newblock \doi{10.1103/PhysRevD.90.025026}

\bibitem{PRD.2011.83.125025}
H.~Belich, E.O. Silva, M.M. Ferreira~Jr., M.T.D. Orlando, Phys. Rev. D
  \textbf{83}(12), 125025 (2011).
\newblock \doi{10.1103/PhysRevD.83.125025}

\bibitem{JMP.2011.52.063505}
K.~Bakke, H.~Belich, E.O. Silva, J. Math. Phys. \textbf{52}(6), 063505 (2011).
\newblock \doi{10.1063/1.3597230}

\bibitem{PRD.2010.82.105023}
K.~Hashimoto, N.~Iizuka, Phys. Rev. D \textbf{82}(10), 105023 (2010).
\newblock \doi{10.1103/PhysRevD.82.105023}

\bibitem{PLB.2011.701.485}
S.R. Dolan, E.S. Oliveira, L.C. Crispino, Phys. Lett. B \textbf{701}(4), 485
  (2011).
\newblock \doi{http://dx.doi.org/10.1016/j.physletb.2011.06.013}

\bibitem{PRL.2010.105.036402}
A.P. Dmitriev, I.V. Gornyi, V.Y. Kachorovskii, D.G. Polyakov, Phys. Rev. Lett.
  \textbf{105}(3), 036402 (2010).
\newblock \doi{10.1103/PhysRevLett.105.036402}

\bibitem{PRL.2012.108.106603}
J.~Schelter, B.~Trauzettel, P.~Recher, Phys. Rev. Lett. \textbf{108}(10),
  106603 (2012).
\newblock \doi{10.1103/PhysRevLett.108.106603}

\bibitem{PRA.2010.82.022104}
A.~Tanaka, T.~Cheon, Phys. Rev. A \textbf{82}(2), 022104 (2010).
\newblock \doi{10.1103/PhysRevA.82.022104}

\bibitem{PLA.2011.375.2952}
D.~Baltateanu, Phys. Lett. A \textbf{375}(32), 2952 (2011).
\newblock \doi{10.1016/j.physleta.2011.06.033}

\bibitem{JMP.2012.53.023514}
K.~Bakke, C.~Furtado, J. Math. Phys. \textbf{53}, 023514 (2012).
\newblock \doi{http://dx.doi.org/10.1063/1.3687022}

\bibitem{PRD.2013.87.125015}
M.A. Anacleto, F.A. Brito, E.~Passos, Phys. Rev. D \textbf{87}(12), 125015
  (2013).
\newblock \doi{10.1103/PhysRevD.87.125015}

\bibitem{PRD.2011.84.045002}
A.~Das, H.~Falomir, M.~Nieto, J.~Gamboa, F.~M\'endez, Phys. Rev. D \textbf{84},
  045002 (2011).
\newblock \doi{10.1103/PhysRevD.84.045002}

\bibitem{PLB.2002.527.149}
M.~Chaichian, P.~Prešnajder, M.~Sheikh-Jabbari, A.~Tureanu, Phys. Lett. B
  \textbf{527}(1–2), 149  (2002).
\newblock \doi{http://dx.doi.org/10.1016/S0370-2693(02)01176-0}

\bibitem{EPJC.2006.46.825}
K.~Li, S.~Dulat, Eur. Phys. J. C \textbf{46}(3), 825 (2006).
\newblock \doi{10.1140/epjc/s2006-02538-2}

\bibitem{PRD.1997.55.5951}
M.L. Horner, A.S. Goldhaber, Phys. Rev. D \textbf{55}(10), 5951 (1997).
\newblock \doi{10.1103/PhysRevD.55.5951}

\bibitem{PRD.1990.42.3524}
C.R. Hagen, S.~Ramaswamy, Phys. Rev. D \textbf{42}(10), 3524 (1990).
\newblock \doi{10.1103/PhysRevD.42.3524}

\bibitem{Book.1975.Reed.II}
M.~Reed, B.~Simon, \emph{Methods of Modern Mathematical Physics. II. Fourier
  Analysis, Self-Adjointness.} (Academic Press, New York - London, 1975)

\bibitem{Book.2004.Albeverio}
S.~Albeverio, F.~Gesztesy, R.~Hoegh-Krohn, H.~Holden, \emph{Solvable Models in
  Quantum Mechanics}, 2nd edn. (AMS Chelsea Publishing, Providence, RI, 2004)

\bibitem{Petiau1936}
G.~Petiau, Published in Acad. Roy. de Belg., Classe Sci., Mem in 8o
  \textbf{16}, 2 (1936)

\bibitem{Kemmer1938}
N.~Kemmer, Proc. R. Soc. Lond. A \textbf{166}, 127 (1938).
\newblock \doi{10.1098/rspa.1938.0084}

\bibitem{PR54:1114:1938}
R.J. Duffin, Phys. Rev. \textbf{54}, 1114 (1938).
\newblock \doi{10.1103/PhysRev.54.1114}

\bibitem{Kemmer1939}
N.~Kemmer, Proc. R. Soc. Lond. A \textbf{173}, 91 (1939).
\newblock \doi{10.1098/rspa.1939.0131}

\bibitem{PLA244:329:1998}
M.~Nowakowski, Phys. Lett. A \textbf{244}, 329 (1998).
\newblock \doi{http://dx.doi.org/10.1016/S0375-9601(98)00365-X}

\bibitem{PLA268:165:2000}
J.T. Lunardi, B.M. Pimentel, R.G. Teixeira, J.S. Valverde, Phys. Lett. A
  \textbf{268}, 165 (2000).
\newblock \doi{http://dx.doi.org/10.1016/S0375-9601(00)00163-8}

\bibitem{PRA90:022101:2014}
L.B. Castro, A.S. de~Castro, Phys. Rev. A \textbf{90}, 022101 (2014).
\newblock \doi{10.1103/PhysRevA.90.022101}

\bibitem{PRD15:1518:1977}
R.F. Guertin, T.L. Wilson, Phys. Rev. D \textbf{15}, 1518 (1977).
\newblock \doi{10.1103/PhysRevD.15.1518}

\bibitem{JPA12:665:1979}
B.~Vijayalakshmi, M.~Seetharaman, P.M. Mathews, J. Phys. A: Math. and Gen.
  \textbf{12}, 665 (1979).
\newblock \doi{10.1088/0305-4470/12/5/015}

\bibitem{PRL26:1200:1971}
E.~Fischbach, F.~Iachello, A.~Lande, M.M. Nieto, C.K. Scott, Phys. Rev. Lett.
  \textbf{26}, 1200 (1971).
\newblock \doi{10.1103/PhysRevLett.26.1200}.
\newblock \urlprefix\url{http://link.aps.org/doi/10.1103/PhysRevLett.26.1200}

\bibitem{PTP51:1585:1974}
E.~Fischbach, M.M. Nieto, C.K. Scott, Prog. Theor. Phys. \textbf{51}, 1585
  (1974).
\newblock \doi{10.1143/PTP.51.1585}.
\newblock
  \urlprefix\url{http://ptp.oxfordjournals.org/content/51/5/1585.abstract}

\bibitem{PRC34:2244:1986}
E.~Friedman, G.~K\"albermann, C.J. Batty, Phys. Rev. C \textbf{34}, 2244
  (1986).
\newblock \doi{10.1103/PhysRevC.34.2244}.
\newblock \urlprefix\url{http://link.aps.org/doi/10.1103/PhysRevC.34.2244}

\bibitem{EPJC.2012.72.2217}
H.~Hassanabadi, Z.~Molaee, S.~Zarrinkamar, Eur. Phys. J. C \textbf{72}, 2217
  (2012).
\newblock \doi{10.1140/epjc/s10052-012-2217-5}

\bibitem{EPJP.2014.129.246}
R.M.T. Cavalcanti, J.M. Hoff~da Silva, R.a. da~Rocha, Eur. Phys. J. Plus
  \textbf{129}(11), 246 (2014).
\newblock \doi{10.1140/epjp/i2014-14246-4}.
\newblock \urlprefix\url{http://dx.doi.org/10.1140/epjp/i2014-14246-4}

\bibitem{PLA316:33:2003}
R.~Casana, V.Y. Fainberg, B.M. Pimentel, J.S. Valverde, Phys. Lett. A
  \textbf{316}(1), 33 (2003).
\newblock \doi{http://dx.doi.org/10.1016/S0375-9601(03)01018-1}.
\newblock
  \urlprefix\url{http://www.sciencedirect.com/science/article/pii/S0375960103010181}

\bibitem{PA419:612:2015}
L.M. Abreu, A.L. Gadelha, B.M. Pimentel, E.S. Santos, Physica A: Statistical
  Mechanics and its Applications \textbf{419}(0), 612 (2015).
\newblock \doi{http://dx.doi.org/10.1016/j.physa.2014.10.049}.
\newblock
  \urlprefix\url{http://www.sciencedirect.com/science/article/pii/S0378437114008875}

\bibitem{EPJC.2015.75.287}
L.B. Castro, Eur. Phys. J. C \textbf{75}(6), 287 (2015).
\newblock \doi{10.1140/epjc/s10052-015-3507-5}

\bibitem{ADHEP2015:901675:2015}
Z.~Wang, Z.W. Long, C.Y. Long, W.~Zhang, AdHEP \textbf{2015}, Article ID 901675
  (2015).
\newblock \doi{http://dx.doi.org/10.1155/2015/901675}.
\newblock \urlprefix\url{http://www.hindawi.com/journals/ahep/2015/901675/}

\bibitem{PRB92:235106:2015}
G.~Palumbo, K.~Meichanetzidis, Phys. Rev. B \textbf{92}, 235106 (2015).
\newblock \doi{10.1103/PhysRevB.92.235106}.
\newblock \urlprefix\url{http://link.aps.org/doi/10.1103/PhysRevB.92.235106}

\bibitem{EPJC76:61:2016}
L.B. Castro, Eur. Phys. J. C \textbf{76}(2), 61 (2016).
\newblock \doi{10.1140/epjc/s10052-016-3904-4}.
\newblock \urlprefix\url{http://dx.doi.org/10.1140/epjc/s10052-016-3904-4}

\bibitem{PRD10:4049:1974}
R.A. Krajcik, M.M. Nieto, Phys. Rev. D \textbf{10}, 4049 (1974).
\newblock \doi{10.1103/PhysRevD.10.4049}

\bibitem{JPA43:055306:2010}
T.R. Cardoso, L.B. Castro, A.S. de~Castro, J. Phys. A: Math. and Theor.
  \textbf{43}, 055306 (2010).
\newblock \doi{10.1088/1751-8113/43/5/055306}

\bibitem{CJP87:857:2009}
T.R. Cardoso, L.B. Castro, A.S. de~Castro, Can. J. Phys. \textbf{87}, 857
  (2009).
\newblock \doi{10.1139/P09-054}

\bibitem{JPA45:075302:2012}
T.R. Cardoso, L.B. Castro, A.S. de~Castro, J. Phys. A: Math. and Theor.
  \textbf{45}, 075302 (2012).
\newblock \doi{10.1088/1751-8113/45/7/075302}

\bibitem{ADHEP2014:784072:2014}
L.B. Castro, L.P. de~Oliveira, AdHEP \textbf{2014}, Article ID 784072 (2014).
\newblock \doi{http://dx.doi.org/10.1155/2014/784072}

\bibitem{UMEZAWA1956}
H.~Umezawa, \emph{Quantum Field Theory} (North - Holland, Amsterdam, 1956)

\bibitem{JPA43:495402:2010}
L.M. Abreu, E.S. Santos, J.D.M. Vianna, Journal of Physics A: Mathematical and
  Theoretical \textbf{43}(49), 495402 (2010)

\bibitem{AoP.2013.339.510}
F.M. Andrade, E.O. Silva, M.~Pereira, Ann. Phys. (N.Y.) \textbf{339}(0), 510
  (2013).
\newblock \doi{10.1016/j.aop.2013.10.001}

\bibitem{PRD.1994.50.7715}
D.K. Park, J.G. Oh, Phys. Rev. D \textbf{50}(12), 7715 (1994).
\newblock \doi{10.1103/PhysRevD.50.7715}

\bibitem{PRD.2012.85.041701}
F.M. Andrade, E.O. Silva, M.~Pereira, Phys. Rev. D \textbf{85}(4), 041701(R)
  (2012).
\newblock \doi{10.1103/PhysRevD.85.041701}

\bibitem{crll.1987.380.87}
F.~Gesztesy, S.~Albeverio, R.~Hoegh-Krohn, H.~Holden, J. Reine Angew. Math.
  \textbf{380}(380), 87 (1987).
\newblock \doi{10.1515/crll.1987.380.87}

\bibitem{JMP.1998.39.47}
L.~Dabrowski, P.~Stovicek, J. Math. Phys. \textbf{39}(1), 47 (1998).
\newblock \doi{10.1063/1.532307}

\bibitem{LMP.1998.43.43}
R.~Adami, A.~Teta, Lett. Math. Phys. \textbf{43}(1), 43 (1998).
\newblock \doi{10.1023/A:1007330512611}

\bibitem{CMP.1991.139.103}
B.S. Kay, U.M. Studer, Commun. Math. Phys. \textbf{139}(1), 103 (1991).
\newblock \doi{10.1007/BF02102731}

\bibitem{JMP.1985.26.2520}
W.~Bulla, F.~Gesztesy, J. Math. Phys. \textbf{26}(10), 2520 (1985).
\newblock \doi{10.1063/1.526768}

\end{thebibliography}

\end{document}